# Bayesian Design of Experiments: Implementation, Validation and Application to Chemical Kinetics

Eric A. Walker[*], Kishore Ravisankar

*Institute for Computational and Data Sciences, State University of New York at Buffalo*

*ericwalk@buffalo.edu (716) 645 8995

**Abstract**

Bayesian experimental design (BED) is a tool for guiding experiments founded on the principle of expected information gain. I.e., which experiment design will inform the most about the model can be predicted before experiments in a laboratory are conducted. BED is also useful when specific physical questions arise from the model which are answered from certain experiments but not from other experiments. BED can take two forms, and these two forms are expressed in three example models in this work. The first example takes the form of a Bayesian linear regression, but also this example is a benchmark for checking numerical and analytical solutions. One of two parameters is an estimator of the synthetic experimental data, and the BED task is choosing among which of the two parameters to inform (limited experimental observability). The second example is a chemical reaction model with a parameter space of informed reaction free energy and temperature. The temperature is an independent experimental design variable explored for information gain. The second and third examples are of the form of adjusting an independent variable in the experimental setup. The third example is a catalytic membrane reactor similar to a plug-flow reactor. For this example, a grid search over the independent variables, temperature and volume, for the greatest information gain is conducted. Also, maximum information gain is conducted is optimized with two algorithms: the differential evolution algorithm and steepest ascent, both of which benefitted in terms of initial guess from the grid search.

**Highlights**

- The information gain metric is obtained through Bayesian inference using a Metropolis-Hastings Markov Chain Monte Carlo algorithm.
- The algorithm is validated against the analytical solution of a Bayesian linear regression.
- For a chemical reaction model, the effect of the design variable temperature is demonstrated on the information gain and the reduction of uncertainty.
- A catalytic membrane reactor model undergoes optimal design of experiments for the temperature and volume with the greatest expected information gain.

## 1. Introduction

Bayesian inference is a powerful tool for updating model uncertainties given experimental data and for model comparison. It has been applied to chemical kinetics[1-8] models and other domain sciences.[9-12] Bayesian inference also has a powerful evidence factor which is a basis for model selection. [9,13] This Bayes' factor consists of a goodness-of-fit term to data and an information gain term (which is a subtracted penalty). Within the Bayesian inference is built-in balance of these two terms. Bayes' factor has been demonstrated to provide a formal metric for model selection, or choosing between two models when given prior estimates including uncertainty for parameters and experimental data.[1] The information gain term provides yet another amount of information and is the metric upon which Bayesian Design of Experiments (BED) is founded. The information gain term is a formalization of Occam's razor that prevents over-fitting the model to the data. In a more strict sense, is a measure of how much changed in the model parameters from a Bayesian inference with experimental data. Therefore, the more the model changed when experimental data became available, the greater the information gain. In a chemical kinetics problem, certain reaction mechanisms and rate-determining steps are ruled out during the Bayesian inference with experimental data. I.e., uncertainty is reduced. The information gain differs at different experimental conditions, and therefore obtaining its value prior to conducting the experiments will result in experiments that answer the most questions raised by the uncertainty in the model.

Uncertainty quantification has recently emerged and spread in the field of heterogeneous catalysis[1,14-22] and homogeneous catalysis[23,24]. A natural extension of the progress of the field will be the design or selecting experimental conditions suggested from theory at which to perform experiments. Materials screening has existed for some time, but experimental design with the objective of gaining depth of knowledge about a select few materials will become a practice. This is despite the underlying principles of BED being in place for a longer time.[25-29] Computational improvements have boosted Bayesian methods in recent years.[25,26] In addition, this is despite Bayesian methods being applied to chemical kinetics,[2,3] though without physics-based first-principles calculations populating these models. All models and variables upon Bayesian statistical tools operate include a specific probability density function (pdf) of the uncertainty. This measure or quantification of uncertainty reflects the realities of predictive physical models is necessary to tabulate when making quantitative predictions about chemical kinetics, for instance asserting the relative dominance of one reaction mechanism over another.[15]

Regarding the application of Bayesian statistical tools in chemical kinetics models, Najm, et al.[3] discussed the use of Bayesian inferences for uncertainty quantification of parameters of a methane-air system ($CH_4 + 2O_2 \leftrightarrow CO_2 + 2H_2O$). The posterior distribution of the parameter space probability density function is inferred by an adaptive Metropolis-Hastings Markov Chain Monte Carlo algorithm, similar to the algorithm employed in this work. Although, their adaptive algorithm overcame numerical challenges in their highly-correlated parameters which were the energies within Arrhenius expressions for rate constants. Inference was performed on synthetic data with added noise. However, the design of experiments to maximize information gain was not the aim of that work. Frenklach, et al.[5] also investigated Bayesian inference for a methane-air combustion model (chemical kinetics). Their model contained a large 102 parameter space, a larger parameter space than Najm, et al..[3] Their model further consisted of 76 quantities of interest (QoI) for which experimental data were available for Bayesian inference. Bayesian inference was compared against a method known as bound-to-bound data collaboration for updating model uncertainties. The authors noted that model bias is inevitable to be encountered with bias being discrepancy variable between model predictions of QoI's experimental observations of the same QoI's. When a term was introduced and Bayesian inferences conducted, the predictions on the QoI's became poorer. Galagali, et al.[8] presented a Bayesian framework in which many possible prior estimates of the parameters are simulated. Their application was the inference of chemical kinetic models of catalytic steam and dry reforming of methane. Reforming of methane is different from combustion, and the product is syngas ($CO + CO_2 + H_2$). Steam reforming includes water as a reactant and dry reforming does not. The aim was to demonstrate computationally efficient comparison of many Bayesian models to select the most optimal model.

Najm, et al.,[6] in a subsequent study, introduced a data-free (although not strictly data free) inference procedure to determine the uncertain parameters in the same methane-air system. There was some knowledge about the bounds (maximums and minimums) of the experiments. This approach is helpful in scenarios where the data required to describe Bayesian frameworks are either insufficient or missing. The presented solution was to generate simulated data (with noise) for the methane-air ignition problem, which is similarly carried out in this work. Using the given data, a chemical model was calibrated with their adaptive[3] Markov Chain Monte Carlo (MCMC) algorithm, which then provided a reference posterior for the parameters of the reaction. The Data-Free Inference (DFI) algorithm made use of the MCMC procedure twice, in the form of an outer and inner chain. The nested MCMC algorithm allowed for exploring the experimental space and parameter space. Though computationally more intensive, with little available information of experiments (their maximum and minimum bounds), a posterior model uncertainty was converged. Kalyanaraman, et al.[7] studied the uncertain parameters in adsorption of $CO_2$ on amine sorbents. Their work most closely resembles this work because the information gain surface was searched for the maximum with an optimization algorithm. The authors point out many of the same principles used in this work including parametric uncertainty versus model uncertainty in Bayesian inference. For information gain the Kullback-Leibler divergence[30] was utilized. After introducing a Bayesian framework, an adaptive Metropolis Hastings Markov Chain Monte Carlo (MCMC) algorithm inferred adsorption isotherm parameters. The information gain was plotted with the interpretation with which the authors of this work coincide, that information gain is the maximal reduction of uncertainty.

Often, experimental data can be scarce. Bayesian design of experiments provides a guiding capability. It is a continuation of expert-based design of experiments. An illustrative example of expert-based design of experiments in chemical kinetics is using in-situ spectroscopy to detect reaction intermediates during a catalytic reaction.[31] Experimental conditions and observations were chosen to answer a proposed physical mechanism. In this work, we demonstrate with an illustrative model of two parameters and synthetic experimental data. As a foundational hypothesis, information gain is assumed to differ depending upon which variables are observed (first type) and also the experimental conditions or independent variables (second type). This hypothesis is strictly true in the first two examples, but not practically true for the agnostic case of the third example. I.e., when synthetic data is entirely unbiased based upon the prior model, the information gain does not change much (chemical kinetics in a reactor example). Therefore, the scenario that the model contains systematic bias or errors depending upon independent variables (temperature and volume) is demonstrated. Specifically, this bias is created by always creating synthetic experimental data from rate constant parameters that are each 10% greater than the prior model expectation. The effect of this parameter discrepancy is expected to cause a smooth divergence in exit flow rates (concentrations) of chemical species between the model prediction and the synthetic experiments over the design space. Such a systematic bias could be estimated by experimentalists or from literature. In addition, systematic bias in the model will be encountered in any model which is not capturing the physical processes underlying experiments, which is bound to occur. In this work, synthetic experimental data generation is part of the process to design experiments since information gain requires a Bayesian inference of the parameters from the model and experiments. At each synthetic experimental design point a Bayesian inference is conducted and the information gain outputted. Each Bayesian inference is a computationally expensive step involving numerical sampling methods in real applications (the first demonstrative example in this work is less complex and may be validated analytically and therefore it serves as a benchmark). The experimental synthetic data may be generated in a grid fashion, at constant intervals, over the plausible maximum and minimum of the design space. Then, each synthetic experimental grid point is inferred for information gain. This approach is demonstrated for the third example, the catalytic membrane reactor. Another method which can prove advantageous in a large-dimensional experimental design space is the use of an optimization algorithm, be it steepest ascent or differential evolution. The principle reason is less Bayesian inferences are required when the maximum information gain is sought optimally rather than through a brute force, grid, search. The catalytic membrane reactor example is evaluated for information gain using these two methods. The steepest ascent method uses numerical gradients as it steps to move towards the maximum information gain over the experimental design space. The differential evolution[32] method optimizes for information gain through series of candidate locations on the surface created by mixing. Gradients are not evaluated, unlike the steepest ascent method.

The two software in this work, both open source, are the Quantification of Uncertainty for Estimation, Simulation and Optimization (QUESO)[33,34] in C++ and the University at Buffalo Random Optimization Algorithm for Models (BuffaloROAM), in Python. BuffaloROAM is introduced for the first time in this work. The mystic framework for Python, that contains various optimization algorithms and tools are used for the differential evolution algorithm for searching for highest information gain.[35,36] In section 2.1, the mathematical and computational framework for experimental design in continuous space is established. Subsequently, the three example models to be studied are specified in section 2.2. Section 2.2 includes an analytical solution of the posterior parameter probability density function (pdf) of the first Bayesian linear regression example model. Section 2.2 conducts a course grid (1-dimensional line) evaluation of Bayesian linear regression example model for Bayesian model evidence and information gain. A robust validation of the Bayesian model evidence and expected information gain is performed in section 3.1 of the results and discussion. The chemical reaction model is similarly evaluated in a grid (1-dimensional line) search in section 3.2. Section 3.3 includes the grid (a 2-dimensional grid) search of the catalytic membrane reactor model. Section 3.3 continues with describing and carrying out the steepest ascent algorithm and the differential evolution algorithm[32].

## 2.1 Experimental design framework

Knuth, et al.[5] have pointed out the spread of Bayesian evidence and Bayes' factor for model selection to domain sciences. Bayesian inference solves (numerically) Bayes formula,

$$p(\theta|D, M_i) = \frac{p(D|\theta, M_i)p(\theta|M_i)}{p(D|M_i)} \qquad (1)$$

$\theta$ is the parameters, for example $\lambda_0$, the exponential growth factor in the exponential model. $D$ is the experimental data. $M_i$ is the model $i$, and there may be multiple models which are compared on the basis of their evidence. In this work, Bayes' formula is performed in a single inference step. However, Bayes' formula may iteratively applied when data is arriving continually in a time series. This is the case when estimating battery charge and remaining useful life during operation when current and voltage data become continually available online. [18] Equation (1) may be rearranged to

$$p(D|M_i) = p(D|\theta, M_i) \frac{p(\theta|M_i)}{p(\theta|D, M_i)} \qquad (2)$$

$$\ln p(\{y_i\}|M_i) = \ln p(\{y_i\}|\theta, M_i) - \ln \frac{p(\theta|D, M_i)}{p(\theta|M_i)} \qquad (3)$$

Integrating the right-hand side over the parameter space $\theta$ (left-hand side is not a function of $\theta$ except with regards to the right-hand side) leads to:

$$\ln p(D|M_i) = E[\ln p(D|\theta, M_i)] - \int_\theta p(\theta|D, M_i) \ln \frac{p(\theta|D, M_i)}{p(\theta|M_i)} d\theta \qquad (4)$$

The evidence for the model is equal to the log-likelihood (goodness of fit) minus the information gain. The integrals over the parameter space $\theta$ in this work are conducted numerically. Both BuffaloROAM and QUESO[33,34] implement Metropolis Hastings Markov Chain Monte Carlo solvers. QUESO is also capable of the more sophisticated multilevel sampling.[37] Both software use a Metropolis Hastings Markov Chain Monte Carlo (MHMCMC). The algorithm proceeds in the following process. An initial $\theta$ is chosen. A matrix $Q$ with zero mean is added to generate a proposal sample,

$$\theta' = \theta + Q \sim \mathcal{N}\left(\begin{bmatrix} 0 \\ 0 \end{bmatrix}, \Sigma_Q\right) \qquad (5)$$

The covariance of Q is typically smaller than the prior covariance to take small steps and not jump too far over the parameter space. The ratio of the probabilities of the proposal sample and the prior sample is evaluated. If the ratio is greater than 1, the probability of accepting the proposal sample is 1. If the ratio is less than 1, that is the probability of accepting the proposal (accept if a random uniform number between 0 and 1 is less than the probability). Equation (6) expresses this computation.

$$p(\theta') = \begin{cases} 1 & \frac{p(D|\theta')p(\theta')}{p(D|\theta)p(\theta)} > 1 \\ \frac{p(D|\theta')p(\theta')}{p(D|\theta)p(\theta)} & \frac{p(D|\theta')p(\theta')}{p(D|\theta)p(\theta)} < 1 \end{cases} \qquad (6)$$

If the proposal sample $\theta'$ is rejected then the current sample is repeated. Multilevel sampling involves flattening the likelihood function to allow the Markov Chain to migrate to the high probability space before restoring the likelihood to its full form.[37] All examples implemented in QUESO in this work are conducted with multilevel sampling. The expected information gain, which is the last term of equation (4), takes the same form of Kullback-Leibler[15,30] (KL) divergence of the posterior and the prior probability density functions (pdf). The KL divergence is a validation of the information gain that is obtained in the general from by subtracting the log evidence from the expected log likelihood. The method of obtaining the information gain by subtraction therefore relies upon the (log) evidence. The expected log likelihood is available from the samples generated from the inference. The analytical evidence may be obtained for a gaussian prior, a linear model and a gaussian likelihood.

### 2.2 Models for study

### 2.2.1 Bayesian linear regression model

The prior for the model is,

$$\mu_{pri} = \begin{bmatrix} 1 \\ 5 \end{bmatrix}, \Sigma_{pri} = \begin{bmatrix} 1 & 0 \\ 0 & 1 \end{bmatrix} \qquad (7)$$

The subscript 'pri' indicates prior. The mean vector is $\mu$ and the covariance matrix is $\Sigma$ is the covariance matrix. The prior is gaussian. The model is,

$$D = X\theta + \epsilon \sim N(\mu_\epsilon = 0, \sigma_\epsilon = 1) \tag{8}$$

In this model, the parameters are estimators for the experiments, $D$. The X matrix is called the design matrix (treated as a linear regression problem). We will consider the first parameter to be observable but not the second parameter, with a design matrix of,

$$X = [1\ 0] \tag{9}$$

The model inadequacy term, $\epsilon$, is mean zero with a variance $(\sigma_\epsilon^2)$ of 1. The model inadequacy term also defines the likelihood. The likelihood is (assuming we are working with one model),

$$p(D|\theta) = \frac{1}{\sqrt{2\pi}\sigma_\epsilon} \exp\left(-\frac{1}{2} \frac{(D-X\theta)^T(D-X\theta)}{\sigma_\epsilon^2}\right) \tag{10}$$

In the case of standard normal covariance, the standard deviation is 1 for both variables contained in $\theta$. The posterior pdf's are,

$$\Sigma_{post} = \left(\Sigma_{pri}^{-1} + \frac{1}{\sigma_\epsilon^2} X^T X\right)^{-1} \tag{11}$$

$$\mu_{post} = \Sigma_{post}\left(\Sigma_{prior}^{-1}\mu_{pri} + \frac{1}{\sigma_\epsilon^2} X^T D\right) \tag{12}$$

We have only 1 data point in $D$. Plugging in our values yields,

$$\Sigma_{post} = \begin{bmatrix} 0.5 & 0 \\ 0 & 1 \end{bmatrix} = \left(\begin{bmatrix} 1 & 0 \\ 0 & 1 \end{bmatrix} + \frac{1}{1^2}\begin{bmatrix} 1 \\ 0 \end{bmatrix}[1\ 0]\right)^{-1} \tag{13}$$

$$\mu_{post} = \begin{bmatrix} 0.5 \\ 5 \end{bmatrix} = \begin{bmatrix} 0.5 & 0 \\ 0 & 1 \end{bmatrix}\left(\begin{bmatrix} 1 & 0 \\ 0 & 1 \end{bmatrix}\begin{bmatrix} 1 \\ 5 \end{bmatrix} + \frac{1}{1^2}\begin{bmatrix} 1 \\ 0 \end{bmatrix}0\right) \tag{14}$$

The data point, $D$, is set to 0. For validation we will check data points 0,1,2 and 3.

$$D = [0,1,2,3], \Sigma_{post} = \begin{bmatrix} 0.5 & 0 \\ 0 & 1 \end{bmatrix}, \mu_{post} = \begin{bmatrix} 0.5 \\ 5 \end{bmatrix}, \begin{bmatrix} 1 \\ 5 \end{bmatrix}, \begin{bmatrix} 1.5 \\ 5 \end{bmatrix}, \begin{bmatrix} 2 \\ 5 \end{bmatrix} \tag{15}$$

When the design matrix changes to express the second parameter, the analytical solution is,

$$X = [0\ 1], D = [0,1,2,3], \Sigma_{post} = \begin{bmatrix} 1 & 0 \\ 0 & 0.5 \end{bmatrix}, \mu_{post} = \begin{bmatrix} 1 \\ 2.5 \end{bmatrix}, \begin{bmatrix} 1 \\ 3 \end{bmatrix}, \begin{bmatrix} 1 \\ 3.5 \end{bmatrix}, \begin{bmatrix} 1 \\ 4 \end{bmatrix} \tag{16}$$

The posterior means and covariances are validated against QUESO[33,34] and BuffaloROAM in section 3.1.

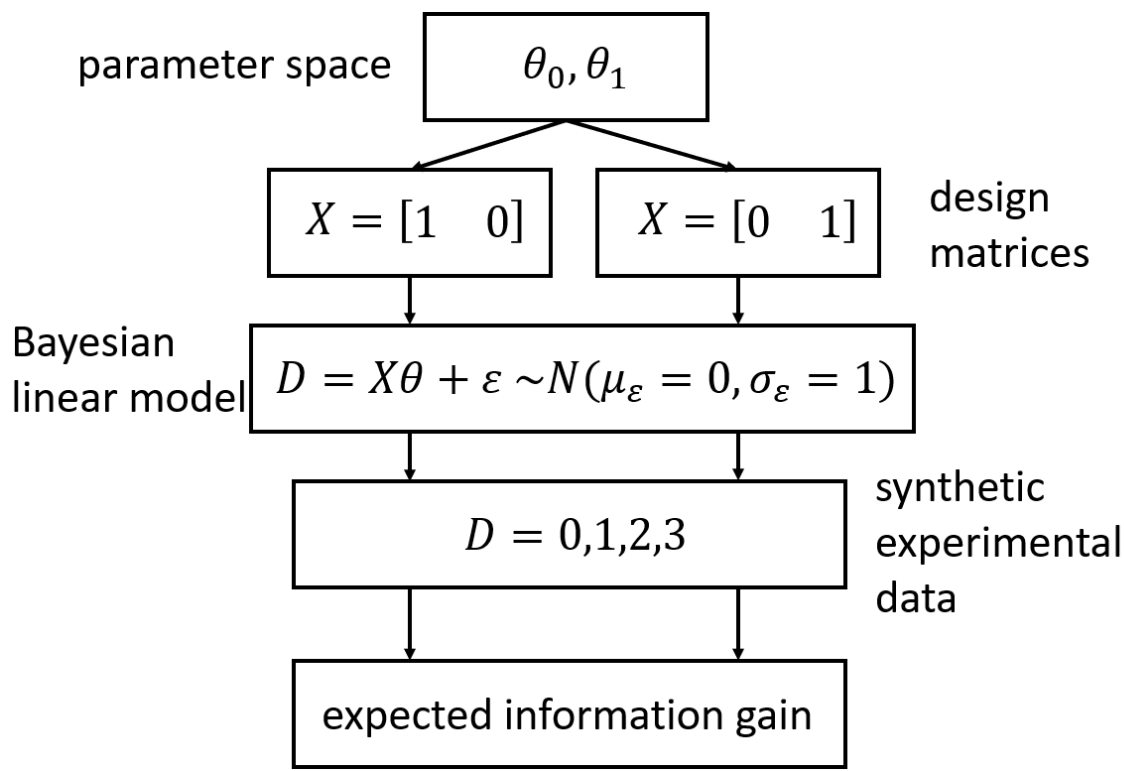

Figure 1. Flow chart of Bayesian linear regression and temperature grid search for chemical reaction model.

### 2.2.2 Chemical reaction model

The next model for examination is a chemical reaction model. This type of model highly non-linear and is of the second class of experimental design problems. I.e., the model has an independent variable, in this case temperature (K), which is not inferred but is part of the experimental conditions. The model describes the reaction,

$$A \leftrightarrow B \tag{17}$$

Equation (17) consists of one reversible reaction. The stoichiometry of the reactant (A) to product (B) is 1:1. The concentration of A is initially set to be 1 prior to reaction. By stoichiometric mass balance, the total concentration of A plus the concentration of B is 1. This remains true when A and B are in a dilute solution without density changes. The concentrations are defined by the mass balance and also an equilibrium constant,

$$C_A + C_B = 1 \tag{18a}$$

$$K = \frac{C_B}{C_A} \tag{18b}$$

Rearranging equations 18a-b and solving for $C_A$ as a function of the equilibrium constant $K$.

$$C_B = K(C_A) \tag{18c}$$

$$C_A + K(C_A) = 1 \tag{18d}$$

$$C_A = \frac{1}{1+K} \tag{18e}$$

$C_A$ and $C_B$ are the concentrations of A and B (unitless), respectively. The equilibrium constant is $K$. Furthermore, the equilibrium constant is a function of the Gibbs free energy by the relation,

$$K = \exp\left(-\frac{\Delta G}{k_B T}\right) \tag{19}$$

$T$ is temperature (K). Boltzmann's constant is $k_B = 8.61733035 \times 10^{-5}$ $(eV K^{-1})$. $\Delta G$ $(eV)$ is the reaction (Gibbs) free energy (eV). The exponential term causes stiff nonlinearity. The expected value of the reaction free energy is -0.2 (eV), corresponding to an exergonic reaction.

The concentration of $C_A$ at equilibrium is the quantity of interest (QoI). Therefore, that concentration goes into the likelihood in the case of Bayesian inference,

$$p(C_A^*|C_A) = \frac{1}{\sqrt{2\pi}\sigma_\epsilon} \exp\left(-\frac{1}{2}\frac{(C_A - C_A^*)^2}{\sigma_\epsilon^2}\right) \tag{20}$$

The asterisk, *, denotes the (synthetic) experimental concentration. The model inadequacy variance is $\sigma_\epsilon^2$. We keep $\sigma_\epsilon^2$ constant for all synthetic experimental data at all temperature points. This is an assumption that the model inadequacy is temperature-independent.

The prior uncertainty pdf of the reaction free energy, $\Delta G$, is,

$$\Delta G_{prior} \sim \mathcal{N}(-0.2, 0.01) \ (eV) \tag{21}$$

The design variable temperature in the denominator influences the spread of the uncertainty in $K$. A lower temperature spreads out the uncertainty in $K$ with the same uncertainty pdf in $\Delta G$ $(eV)$, as long as $\Delta G$ $(eV)$ is negative. Concentration is transformed to log concentration, $\ln(C_A)$, due to the exponential term in equation 19. Five temperature points at 100 (K) intervals between 298.15 and 698.15 (K) are simulated. The expected (value) concentration of $C_A$ increases as the temperature increases as long as $\Delta G$ $(eV)$ remains negative and the reaction is exergonic. The prior probability of a negative reaction free energy is infinitesimal, and therefore the reaction with prior uncertainty is always exergonic. Therefore, by the relation of equation 18, the concentration of A always increases for a particular $\Delta G$ $(eV)$.

### 2.2.3 Catalytic membrane reactor model

A catalytic membrane reactor[38-43] model is introduced. Membrane reactors selectively remove one chemical component from the system with a membrane. This technology drives the equilibrium of the chemical reaction occurring in the reactor in a desired way. This model is introduced because two design parameters are involved, reactor volume and temperature. With two experimental design parameters, the search for the optimal experimental design is more efficiently conducted with steepest ascent than grid search. Catalytic membrane reactors may be modeled with partial differential equations, which may be solved using finite difference or finite element methods[44-47]. In this work we make assumptions that reduce the dimension of the differential equations to one (ordinary differential equations). We will assume again two components as in equation 16. $B$ is assumed to be small enough to pass through the microscopic membrane, but not $A$. A mass balance is conducted,[48] yielding,

$$\frac{dF_A}{dV} = -r_A \tag{22a}$$

$$\frac{dF_B}{dV} = r_A - R_B \tag{22b}$$

$F_A, F_B \left(\frac{mol}{s}\right)$ are the molar flow rates of $A$ and $B$, respectively. $V(cm^3)$ is the reactor volume which is an experimental design parameter. The reaction rate of $A$ to $B$ (reversible) is $r_A$. The flow through the membrane of component $B$ is denoted by $R_B$. The two reactions, $r_A$ and $R_B$ are defined,

$$r_A = -k_1 C_A + k_{-1} C_B \tag{23a}$$

$$R_B = k_m C_B \tag{23b}$$

The rate constant for transport of $B$ through the membrane is $k_m (s^{-1})$. The forward and reverse rate constants are $k_1 (s^{-1})$ and $k_{-1} (s^{-1})$, respectively. $C_A \left(\frac{mol}{cm^3}\right)$ and $C_B \left(\frac{mol}{cm^3}\right)$ are obtained by,

$$C_A = C_{T0} \left(\frac{F_A}{F_T}\right) \tag{24a}$$

$$C_B = C_{T0} \left(\frac{F_B}{F_T}\right) \tag{24b}$$

$C_{T0} \left(\frac{mol}{cm^3}\right)$ is the total initial concentration entering the reactor of both components (vapor phase). $F_T \left(\frac{mol}{s}\right)$ is the total molar flow rate of $A$ and $B$.

$$F_T = F_A + F_B \tag{25}$$

$$X = \frac{F_{A0} - F_A}{F_{A0}} \tag{26}$$

This requires an initial assumption of an initial molar flow rate of $A$ of $F_{A0} = 10 \left(\frac{mol}{s}\right)$. $X$ is the conversion, where 0 is no $A$ reacted and 1 is all $A$ reacted. The initial molar flow rate of $B$ is $F_{B0} = 0 \left(\frac{mol}{s}\right)$. The ideal gas law is assumed, yielding,

$$C_{T0} = \frac{P_0}{R(T_0)} \tag{27}$$

$C_{T0} \left(\frac{mol}{cm^3}\right)$ is the total initial concentration of both $A$ and $B$. $P_0 (atm)$ is the initial pressure (assumed not to change). $R \left(\frac{cm^3 atm}{Kmol}\right)$ is the gas constant and is equal to $82.057338\left(\frac{cm^3 atm}{Kmol}\right)$. The volume $V$ $(cm^3)$ range is set from 0 to 1000 ($cm^3$) in this work. The catalyst is located at the membrane boundary, yet the concentration is assumed consistent throughout the reactor by assuming no mass transport limitations. The reactor is also assumed isothermal and isobaric. All code and data for the work are available at https://bitbucket.org/ericawalk/datascience/src/master/.

## 3. Results and Discussion

### 3.1 Bayesian linear regression and quantity validations

The Bayesian linear regression model was grid-searched for expected information gain. This was a coarse grid search at the synthetic experimental data points 0, 1, 2 and 3, with a variance of $\sigma_\epsilon^2 = 1$ at each point (equation 10). Figures 2-3 plot the prior and posteriors when the experiment is 1 and $X = [1\ 0]$ (figure 2) (b)$X = [0\ 1]$ (figure 3). Figures 2-3 demonstrate the influence of the design matrix $X$, or which of the two parameters observes the experiment. Given the two parameters have different priors, their posterior distributions demonstrate inference when the inference is exactly centered on the experiment and when the prior is far from the experiment. The analytical solution for figure 2, from section 2.2.1 equation 15, is $\Sigma_{post} = \begin{bmatrix} 0.5 & 0 \\ 0 & 1 \end{bmatrix}, \mu_{post} = \begin{bmatrix} 1 \\ 5 \end{bmatrix}$. In figure 2, the prior is over the experiment, and the posterior becomes more certain upon the Bayesian inference step (variance is less). The expected value has not changed, but the uncertainty has become greater as the experimental observation confirmed the expectation value. In figure 3, the prior is far from the experiment, and the posterior is centered halfway between the prior and the experiment, in agreement with the analytical solution $\Sigma_{post} = \begin{bmatrix} 1 & 0 \\ 0 & 0.5 \end{bmatrix}, \mu_{post} = \begin{bmatrix} 1 \\ 3 \end{bmatrix}$. Figure 3 demonstrates the balance struck between prior belief and experimental observation.

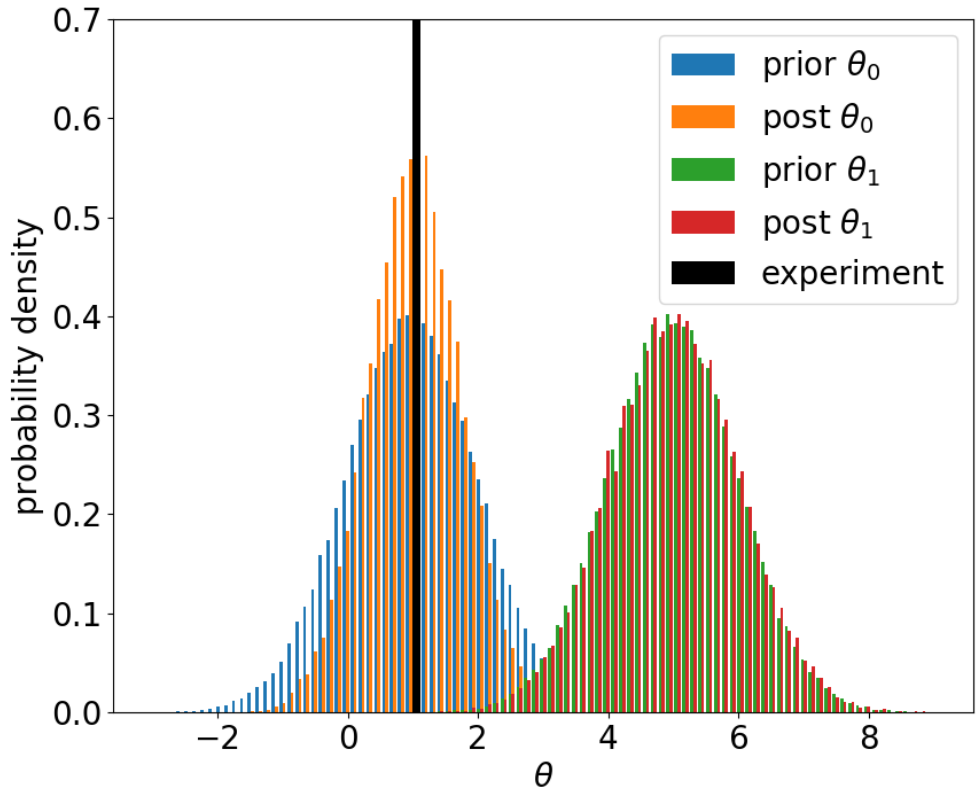

Figure 2. Bayesian linear regression when the design matrix is $X = [1\ 0]$. The second parameter, $\theta_1$ is inferred. The analytical solution from equation 16 is $\Sigma_{post} = \begin{bmatrix} 0.5 & 0 \\ 0 & 1 \end{bmatrix}$, $\mu_{post} = \begin{bmatrix} 1 \\ 5 \end{bmatrix}$.

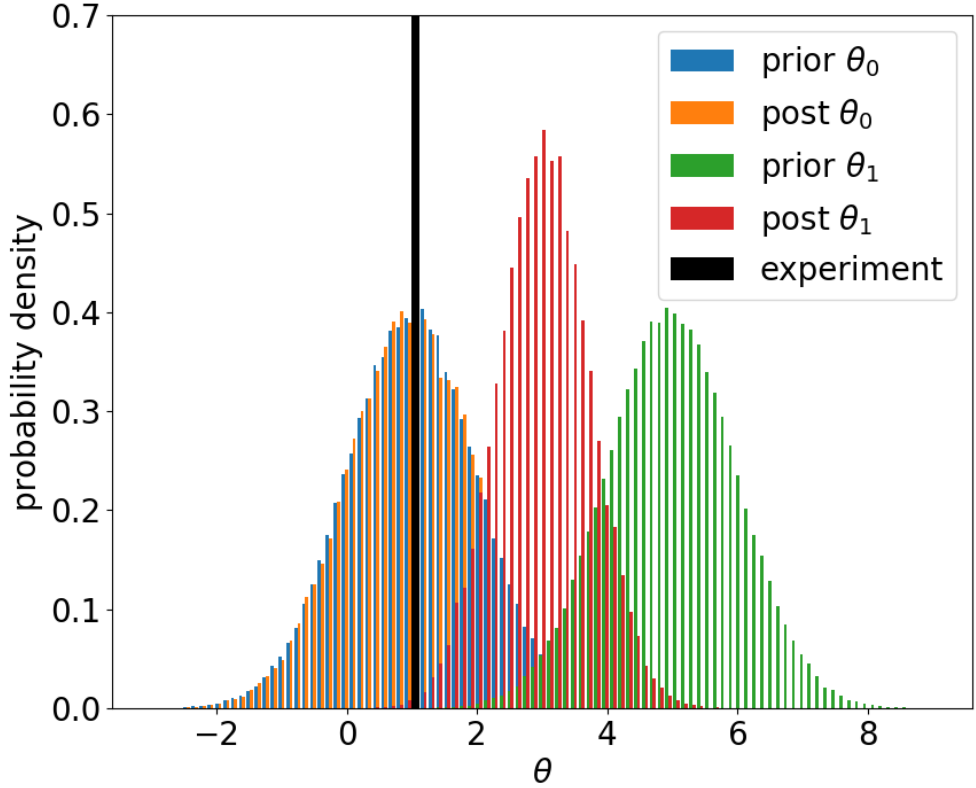

Figure 3. Bayesian linear regression when the design matrix is $X = [0\ 1]$. The second parameter, $\theta_1$ is inferred. The analytical solution from equation 16 is $\Sigma_{post} = \begin{bmatrix} 1 & 0 \\ 0 & 0.5 \end{bmatrix}, \mu_{post} = \begin{bmatrix} 1 \\ 3 \end{bmatrix}$.

The general scenario to obtain information gain is by subtracting the evidence from the expected log likelihood, a rearrangement of equation 4. This direct subtraction method is implemented by QUESO and by BuffaloROAM. Figure 4 shows the expected information gain (from QUESO and BuffaloROAM) for $X = [1\ 0], [0\ 1]$ for experiments 0,1,2 and 3. The two software generate qualitatively the same shape surface. When the design matrix is $X = [1\ 0]$, the minimum information gain is when the experiment is 1. The reason is the prior mean is located at 1 for $\theta_0$. The curve is then symmetrically rising when the experiment is 0 or 2, because the posterior is moving towards those experimental values away from the prior at 1. The most information gain is when the experiment is 3, furthest from the prior of 1. When the design matrix is $X = [0\ 1]$, the information gain tends to be higher, because the prior of 5 is further from experiments. The exception is when the experiment is 3. When the experiment is 3, the distance from the prior values of 1 or 5 corresponding to $\theta_0$ and $\theta_1$, the distance is equal to 2. For that reason, the expected information gains match. The information gain when the design matrix is $X = [0\ 1]$ may be further examined for consistency. The information gain rises as the experiment moves further away from 5, and the information gain is higher than when the design matrix is $X = [1\ 0]$, corresponding to greater distances of the experiment from the prior.

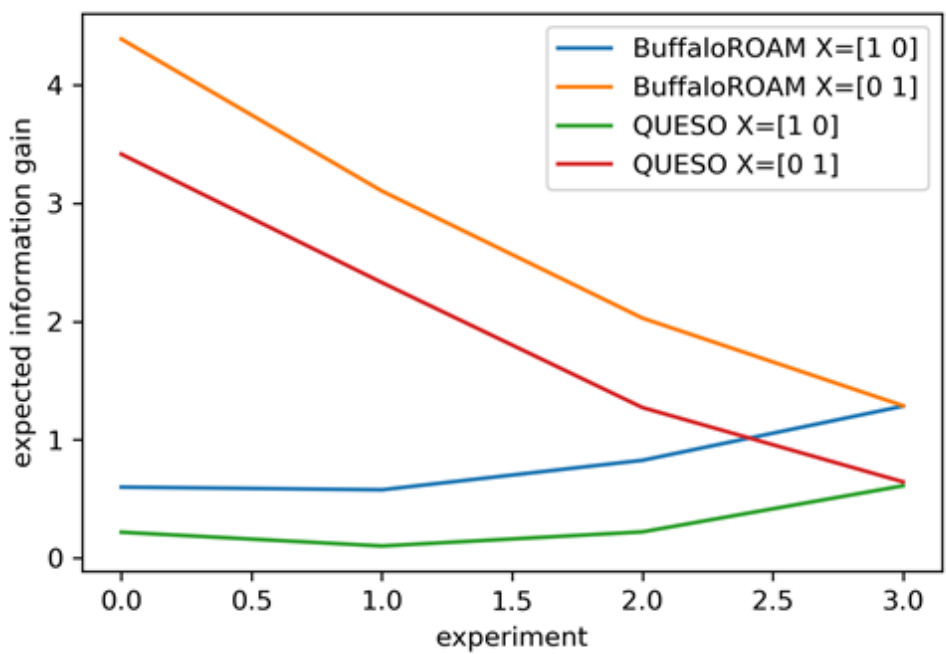

Figure 4. The expected information gain versus the synthetic experiment in QUESO[25,26] and BuffaloROAM (in Python). The two different design matrix scenarios are plotted for each software.

### 3.2 Chemical reaction experimental design

Figure 5 illustrates the grid-search process for design of experiments. At each temperature the information gain is evaluated over the same parameter $\Delta G\ (eV)$. The model is the same over the search except for the independent variable of temperature, $T$. Through the rearrangement of the equilibrium equation, 18b, and the stoichiometry and mass conservation of equation 18a, the concentration of A is obtained from equation 18e. The model discrepancy term is gaussian, zero-mean, signifying no known bias, and the standard deviation is set to a constant 0.025 (dimensionless). Synthetic experiments were generated which assumed the prior $\Delta G$ and model to be exactly accurate. Conceptually, this is similar to Figure 2 and its associated scenario in which the observed variable is exactly the expectation value, and the posterior changes by becoming more certain of the expected value. However, the model is not Bayesian linear regression and is nonlinear. An analytical solution is not tractable and the Metropolis Hastings MCMC algorithm is directly employed. From this search 5 information gain values are obtained, as well as the posterior $\Delta G$'s and concentrations.

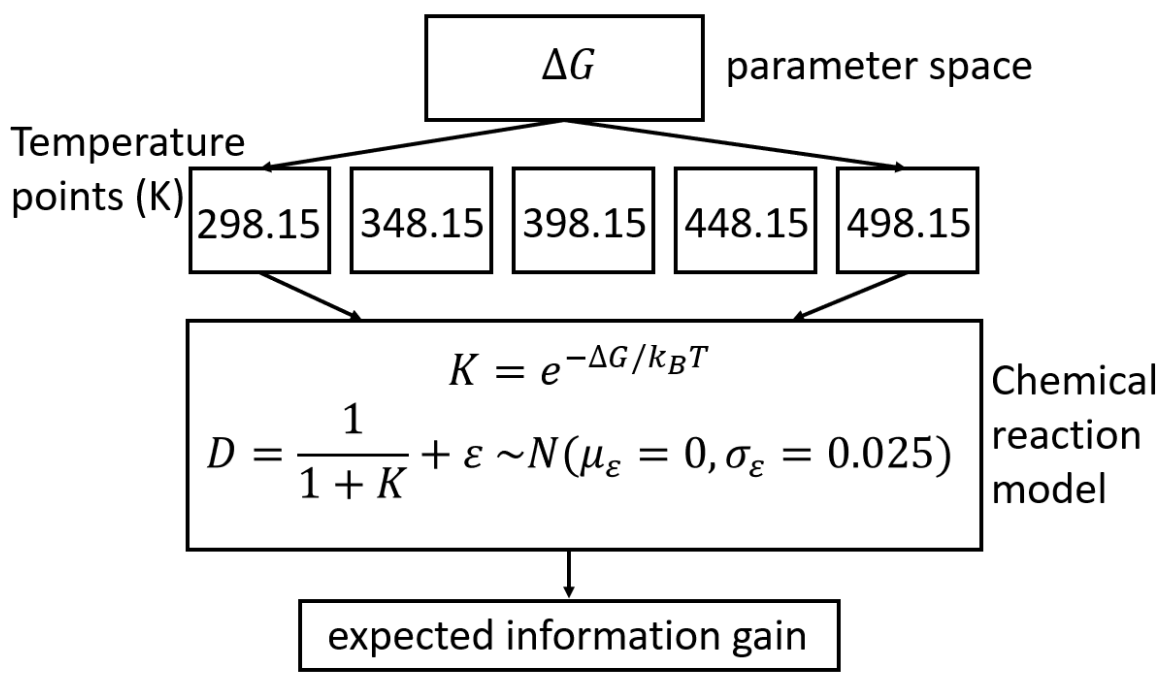

Figure 5. Flowchart of chemical reaction experimental design.

Figure 6(a) displays the results of the process of figure 5. Figure 6a illustrates the prior uncertainty before Bayesian inference in the form of a violin plot where higher probability density is represented by greater width. The violin plot utilizes a kernel density estimator to represent pdf's. A conceptual trend of figure 6(a), which can be seen from equation 19, is that the concentration varies more over the uncertain confidence interval, which is consequential to the information gain. Since $T$ is in the denominator of a positive fraction inside an exponent, a smaller temperature over the same prior $\Delta G$ corresponds to a lesser spread or distribution. A change in expected information gain is expected over the temperature points because while the model discrepancy is unchanging, the prior uncertainty is wider at lower temperatures, and the result is the change in uncertainty of $\Delta G$ is of greater magnitude at lower temperatures. The information gain monotonically decreases with increasing temperature. The computed information gain confirms this trend in Figure 6b. Figure 6c the prior and posterior uncertainty in $\ln(C_A)$ versus temperature in the form of a violin plot. To further verify this, the prior and posterior $\ln(C_A)$ are plotted from 298.15 (K) to 698.15 (K) in figure 6c. The posterior uncertainty is sharper (has a smaller variance) relative to the prior uncertainty, when the temperature is 298.15 (K) than when the temperature is 698.15 (K). This corresponds to a greater information gain about the model uncertainty when the temperature is lower. The prior and posterior $\Delta G$'s are plotted and displayed in the supporting information.

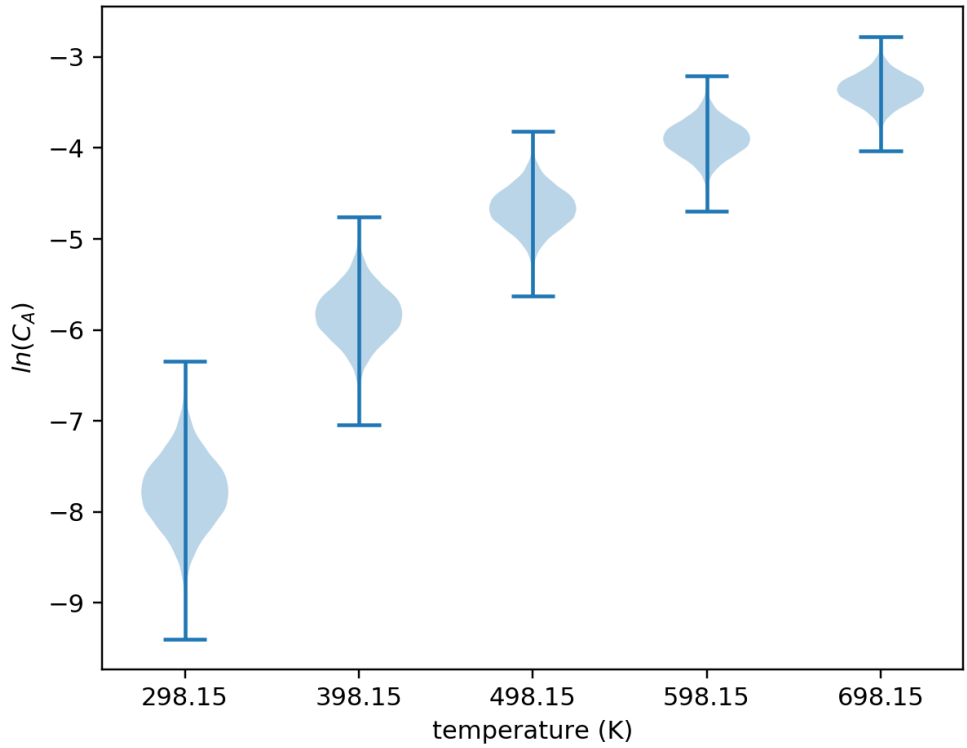

(a)

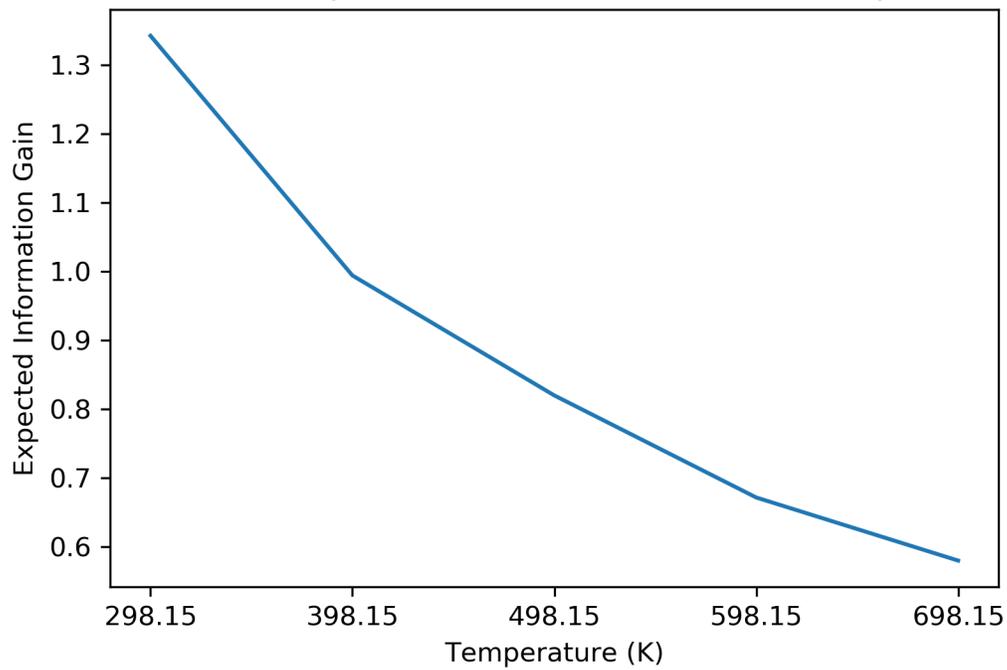

(b)

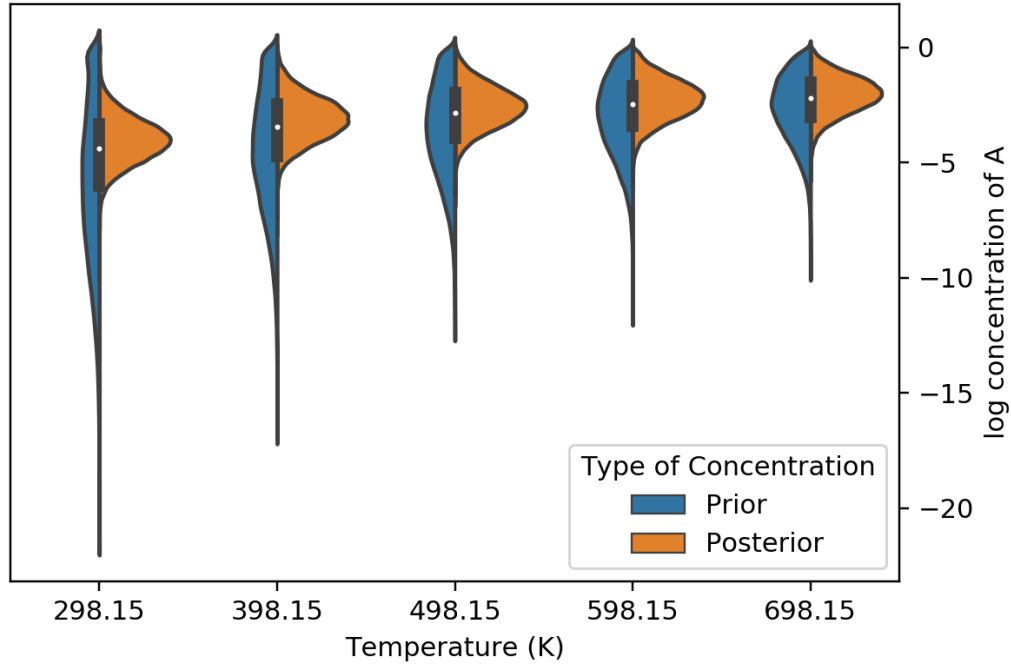

(c)

Figure 6. Chemical reaction design of experiments. (a) Violin plot of the forward uncertainty of the chemical reaction model for the concentration of *A*. (b) Expected information gain versus temperature. (c) prior and posterior uncertainty at 298.15 (K) and 698.15 (K).

### 3.3 Catalytic membrane reactor with optimal experimental design

The membrane reactor model involves three parameters; $k_1$, $k_{-1}$ and $k_m$. Prior pdf's (uncertainties) for these parameters are set,

$$\begin{bmatrix} k_1 \\ k_{-1} \\ k_m \end{bmatrix} \sim N\left(\begin{bmatrix} 1 \times 10^2 \\ 1 \\ 10 \end{bmatrix}, \begin{bmatrix} 10 & 0 & 0 \\ 0 & 10 & 0 \\ 0 & 0 & 10 \end{bmatrix}\right) \tag{28}$$

and an experimental design space is delineated consisting of reactor volume and inlet gas temperature. A schematic of the catalytic membrane reactor described in section 2.2.3 is displayed in figure 7. (a)

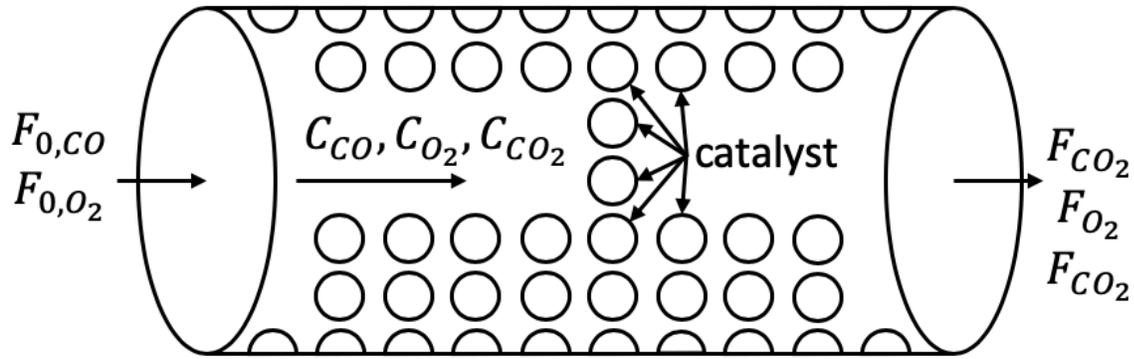

(b)

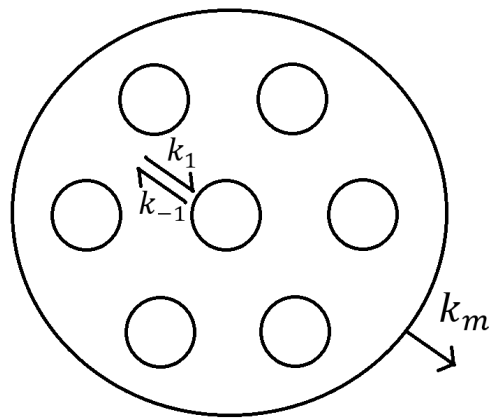

Figure 7. Catalytic plug flow reactor model. (a) reactor diagram (b) Density functional theory calculated reaction intermediates.

The quantity of interest or experimental observable is the outlet flow rate of species A. A general trend that was observed over the design space was that the flow rate of A decreased as the volume of the reactant increased, with the temperature exhibiting a lesser effect. There was a marginal increase in the flow rate of the reactant as the temperature of the reaction increased, irrespective of volume.

The Metropolis Hastings MCMC algorithm is used for each inference at different temperatures and volumes in the search space, and it generates the expected information gain. The temperature is varied from 298.15K to 1000K, and the volume is varied from 100 cm$^3$ to 1100 cm$^3$.

A peak occurs in the search grid when the volume is 1100 cm³, while the temperature is maintained at 398.15K. The information gain at this point is 2.691. The biggest dip in the information gain is seen when the temperature increases beyond 498.15K, and the volume decreases from 1100 cm³ to 350 cm³. The least information gain obtained is 2.492, at 498.15K and 600 cm³. A general trend that can be observed is that the information gain is considerably high when the volume in the reactor is maintained at higher volumes and lower temperatures.

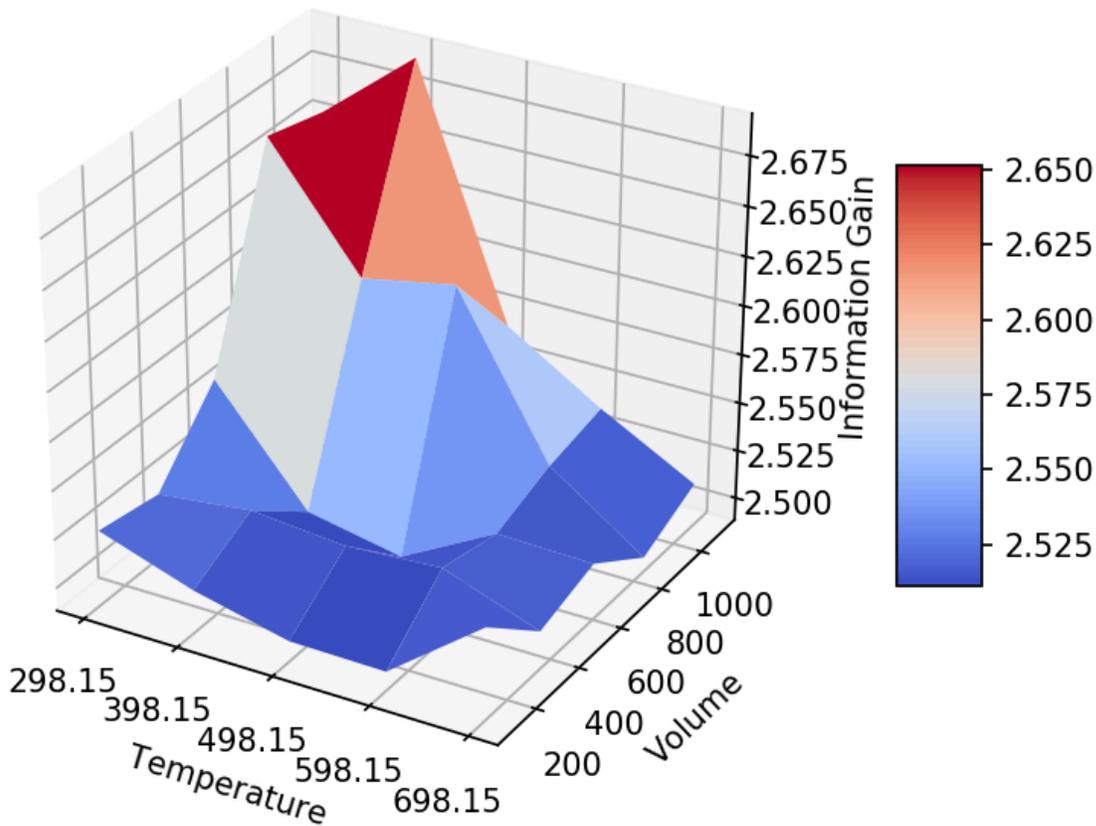

Figure 8. Info gain versus temperature chemical reactor model.

The steepest ascent algorithm is used to maximize the function which yields the expected information gain. The algorithm moves towards the maximum of the function by taking steps towards the positive of the gradient. The steepest ascent algorithm is implemented with alternating steps along the path, in order to obtain gradients with respect to temperature and volume, separately. One of the prime advantages of this method is the faster and simpler computation of the function value. The limitation of this method is the non-convergence when the function is ill-conditioned or non-differentiable. The path of the steepest ascent algorithm is shown in figure 9. A stabilizing constant in the denominator was added to prevent large jumps in the design space. The steepest ascent algorithm relies upon an initial guess of the direction in the space (or on the surface) of the first step to take, as well as a learning rate parameter. The initial guess deployed for the steepest ascent makes use of knowledge of the information gain surface from a grid search, but ideally the grid search is not necessary. The differential evolution algorithm may be used, and it requires a search space and initial guess also, but does not require an initial path direction nor learning rate. The differential evolution algorithm was developed to minimize functions that were non-linear and non-differentiable in nature.[32] It is a derivative free algorithm that can search large spaces to obtain an optimal solution with respect to a given measure. It does not make use of conventional gradient based methods, hence the number of evaluations taken are slightly higher.

The algorithm starts by having initializing some candidate solutions at random positions, in the given search space. Each candidate is represented as a vector that contains some real numbers, that are analogous to the parameters of the given function. The algorithm then works on minimizing the numbers, by computing a new candidate solution by using the existing candidates. This is achieved using the process of mutation and recombination using random distributions. If the parameters of the new solution are found to be better than any of the existing solutions, the old candidate is replaced following the function evaluation with the new candidate. This process is repeated until the termination criterion is met. The criteria could either be the number of iterations of the algorithm, or the tolerance of the solution. Upon execution, the differential evolution algorithm tended to stay close to the intial guess of 450 K and 900 $cm^3$, with a maximum information gain located at 439 K and 890 $cm^3$.

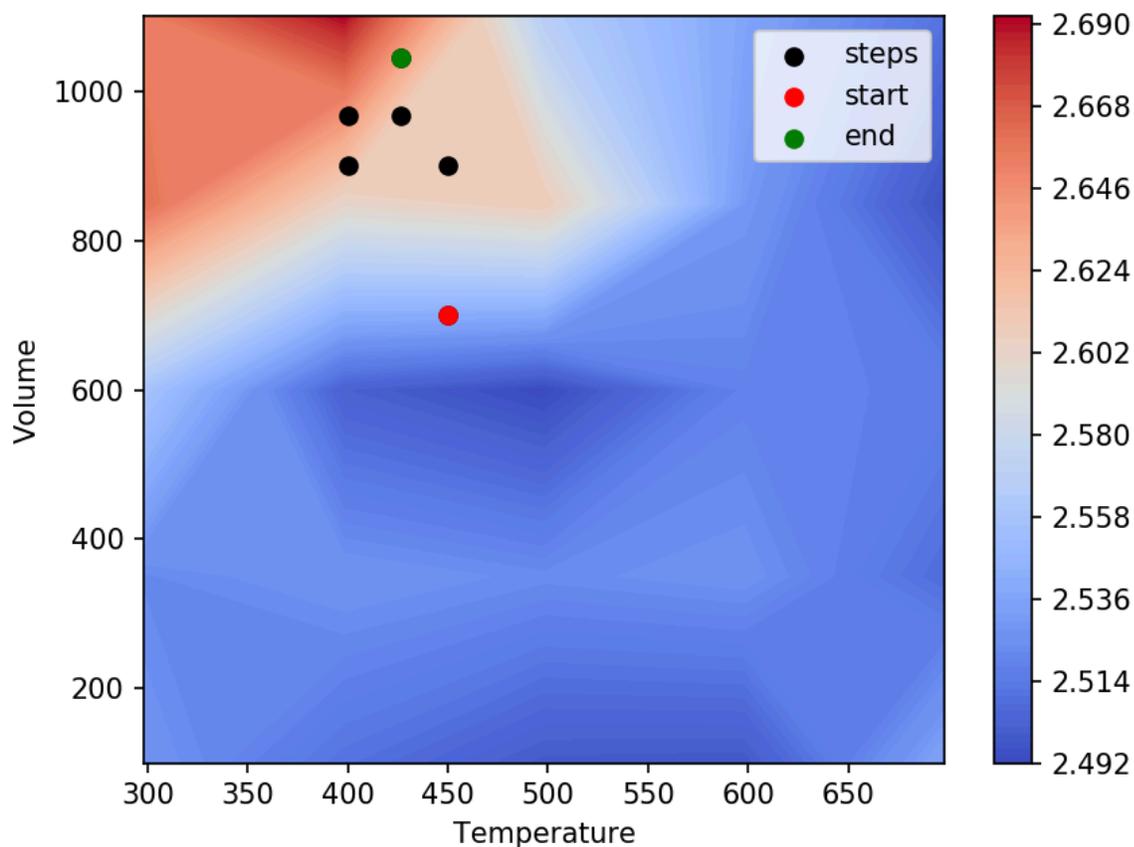

Figure 9. Steepest ascent path over the design space on the information gain surface for the catalytic membrane reactor example.

## 4. Conclusion

Through three demonstrating examples, the principle of experimental design is explored and implemented, in particular for chemical kinetics applications, although in principle for any linear or non-linear model with parameters and quantities of interest. The Bayesian linear regression model served as a benchmark for BuffaloROAM, the software introduced in this work. Both QUESO and analytical evaluation (exact solution known for Bayesian linear regression) validate BuffaloROAM. The chemical reaction experiment design model demonstrated the principle of maximum information gain as it relates to reducing uncertainty maximally. The expected information gain, along the single design dimension of temperature, was significantly high at lower temperatures. This was due to the nature of the model and the Arrhenius expression, that an uncertainty in a negative free energy of reaction would be more spread in concentration at lower temperatures. The experimental design trend included the assumption that model discrepancy with experiments on the quantity of interest, concentration, would be constant across temperature. In that case, the optimal experimental design is at the lowest temperature. The final example was a catalytic membrane reactor model. In this model, a grid search of the design space was conducted, followed by optimal design algorithms. The optimal design algorithms were steepest ascent and differential evolution. Key differences of the two algorithms were that steepest descent is gradient-based and differential evolution makes combinations of guesses without using gradients. Both algorithms benefitted from making initial guesses based upon the brute force grid search. Therefore, a combination of grid search and optimization algorithm is suggested for finding optimal experimental design, especially when the model is non-linear and a rule of thumb for finding highest information gain cannot be readily obtained.

## Acknowledgements

The authors acknowledge the Center for Computational Resources at the State University of New York at Buffalo for computational resources and for the expert support of Ms. Cynthia Cornelius. Dr. Tim Adowski and Prof. Paul Bauman are acknowledged for helping to conceive the problem through numerous discussions and sharing of knowledge.

**Supporting Information: Bayesian Design of Experiments: Implementation, Validation and Application to Chemical Kinetics**

Eric A. Walker, Kishore Ravisankar

This chemical reaction model of section 2.2.2 and figure 6 uses the Metropolis-Hastings Markov Chain Monte Carlo (MCMC) algorithm to conduct Bayesian inferences at five different temperatures: 298.15K, 398.15K, 498.15K, 598.15K and 698.15K. The inferred parameter is the Gibbs free energy, $\Delta G$ $(eV)$. The prior and posterior from the Bayesian inference at each temperature are plotted in figure S1. At T=298.15K, it can be observed that the posterior is much sharper, and the difference between the prior and posterior peak values is quite high, in accordance with the associated quantity of interest, concentration, in figure 6c. With increase in temperature, a lowering in this difference can be observed, and the sharpness of the peak reduces. In other words, the peak posterior value decreases with rise in temperature.

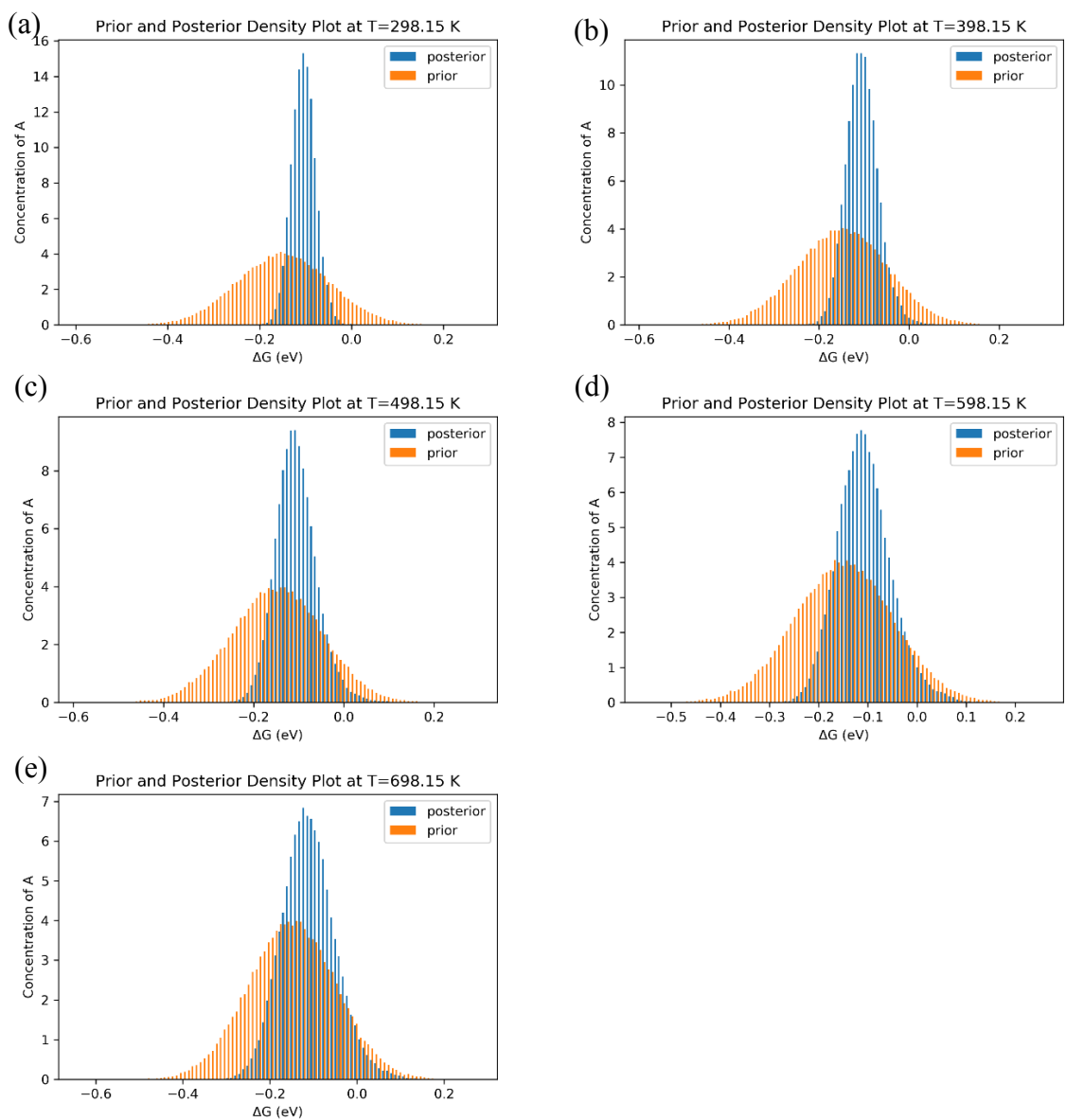

Figure S1. Prior and posterior $\Delta G$ $(eV)$ across temperatures for the chemical reaction model. The posterior is more informed at lower temperatures than higher temperatures.